\documentclass[lettersize,journal]{IEEEtran}

\usepackage{amsmath}
\usepackage{amssymb}
\usepackage{graphicx}
\usepackage{booktabs}
\usepackage{tabularx}
\usepackage{array}
\usepackage{pifont}
\usepackage{multirow}
\usepackage{makecell}
\usepackage[most]{tcolorbox}
\usepackage{enumitem}
\usepackage{xcolor}
\usepackage{url}
\usepackage[hidelinks]{hyperref}

\usepackage{cite}
\usepackage[caption=false,font=footnotesize]{subfig}

\graphicspath{{Figures/}}

\newcommand{\tifsAuthorNames}{Boyu Qiao, Zixin Tang, Yunman Chen, Xiaoshuai Hao, Dongyu She, Wei Zhou, Yunfeng Diao, Songlin Hu}
\newcommand{\tifsShortAuthorNames}{Qiao et al.}
\newcommand{\tifsAffiliationText}{
Boyu Qiao, Yunman Chen, and Songlin Hu are with the Institute of Information Engineering, Chinese Academy of Sciences, Beijing, China, and also with the School of Cyber Security, University of Chinese Academy of Sciences, Beijing, China. E-mail: \{qiaoboyu, husonglin\}@iie.ac.cn, pux754736@gmail.com.
Zixin Tang and Dongyu She are with Zhongguancun Laboratory, Beijing, China. E-mail: \{tangzixin, shedy\}@zgclab.edu.cn.
Xiaoshuai Hao is with Xiaomi EV, Beijing,China. E-mail: \{haoxiaoshuai714\}@163.com. Wei Zhou is with the Institute of Information Engineering, Chinese Academy of Sciences, Beijing, China. E-mail: \{zhouwei\}@iie.ac.cn. Yunfeng Diao is with the School of Computer Science and Information Engineering, Hefei University of Technology, Hefei, China. E-mail: \{diaoyunfeng\}@hfut.edu.cn. (Corresponding authors: Zixin Tang, Wei Zhou.)
}

\begin{document}

\title{Generalizable Social Bot Detection via Multi-Granularity Summarization and Domain-Invariant Learning}

\author{\tifsAuthorNames
\thanks{\tifsAffiliationText}
}

\markboth{arXiv Preprint}
{\tifsShortAuthorNames: MGDIL}

\maketitle

\begin{abstract}
Social bots are becoming increasingly sophisticated in mimicking human users, posing persistent threats to the integrity of online information ecosystems.
Despite substantial progress in social bot detection, existing methods often exhibit limited generalizability when applied to previously unseen datasets.
This limitation mainly arises from two challenges.
First, user information varies substantially in schema and availability across datasets, making it difficult to construct consistent representations from heterogeneous user evidence.
Second, distribution shifts across datasets may cause models trained on source domains to rely on domain-specific patterns that fail to generalize to unseen domains.
To address these challenges, we propose \textbf{Multi-Granularity Summarization and Domain-Invariant Learning (MGDIL)}, a unified framework for generalizable social bot detection.
MGDIL follows a two-stage design that first unifies heterogeneous user evidence and then learns transferable and class-discriminative representations.
It first converts profile attributes into structured textual descriptions and summarizes historical posts into user-level behavioral representations, providing a shared semantic interface across heterogeneous datasets.
Based on the unified representations, MGDIL further combines task-oriented instruction tuning, domain-adversarial learning, and cross-domain contrastive learning to reduce dependence on source-domain-specific patterns while preserving discriminative information between bots and human users.
We construct a unified instruction-tuning corpus from 15 social bot datasets, train MGDIL on 13 earlier datasets, and evaluate it on two later-collected unseen datasets under a temporally ordered cross-dataset protocol.
Experimental comparisons with representative conventional, neural, LLM-based, and transfer-based baselines, together with ablation and representation analyses, demonstrate the effectiveness of MGDIL in improving social bot detection generalization across datasets. The code is available at: https://github.com/QQQQQQBY/MGDIL.
\end{abstract}

\begin{IEEEkeywords}
Social bot detection, cross-domain generalization, domain-invariant learning, instruction tuning.
\end{IEEEkeywords}

\section{Introduction}
Social media platforms have become central to information dissemination, public communication, and opinion formation, but they are increasingly threatened by social bots.
By automating large-scale interactions, imitating legitimate users, and amplifying misleading or manipulative content, social bots can undermine the integrity and trustworthiness of online information ecosystems \cite{cresci2020decade, lopez2025dissecting}.
Recent advances in generative artificial intelligence further enhance their camouflage capabilities, enabling social bots to produce increasingly human-like profiles, language, and interaction patterns \cite{ferrara2023social,qiao2025dynamic}. Detecting social bots is thus critical for safeguarding platform integrity and ensuring trustworthy online communication \cite{feng2024does}.

The problem of social bot detection has received extensive attention from the research community. Early approaches mainly relied on manually designed user metadata \cite{yang2020scalable, hayawi2022deeprobot}, behavioral statistics \cite{cresci2016dna,di2023online}, and textual features \cite{wei2019twitter,dukic2020you}. More recent studies exploit richer user information, including historical posting sequences \cite{qiao2023social,liu2023botmoe}, behavioral patterns \cite{wu2023botshape,he2024dynamicity}, and relation-aware graph structures \cite{li2023multi,feng2022heterogeneity}. Meanwhile, advances in large language models have further strengthened the ability of social bots to mimic human linguistic and behavioral characteristics \cite{qiao2025botsim, liu2025evolution}.
These developments make social bot detection increasingly challenging
when detectors are deployed beyond the datasets on which they are trained. Across datasets, user information can vary substantially in both schema and availability, while differences in bot populations, topical distributions, platform environments, and data collection protocols may further induce significant distribution shifts. Consequently, generalizable social bot detection requires models to construct consistent and informative representations from heterogeneous user evidence while remaining discriminative between bots and human users on previously unseen datasets \cite{song2025self, sun2025adaptive}.

\begin{figure}[t]
  \centering
  \includegraphics[width=\columnwidth]{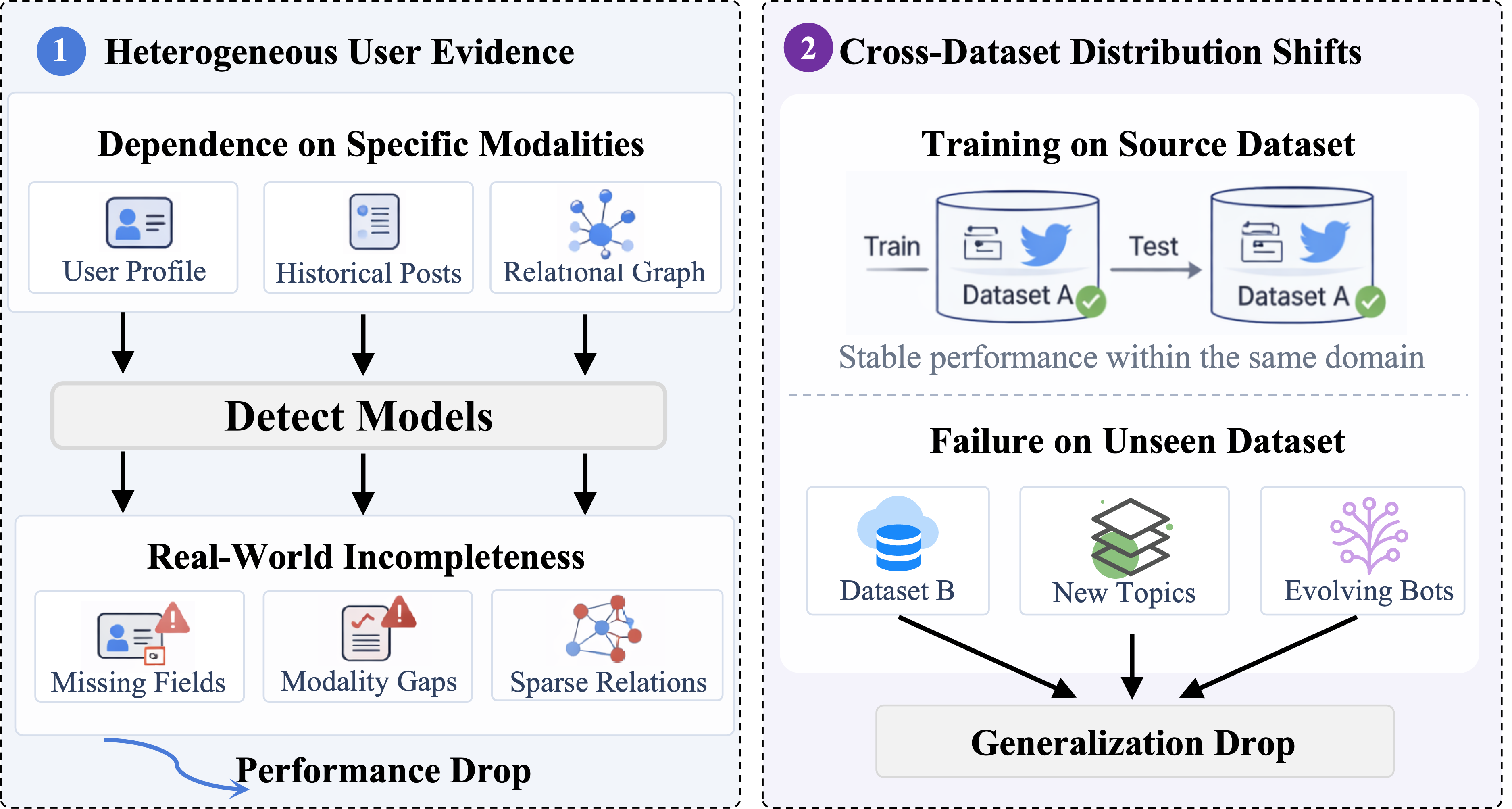}
  \caption{Two challenges in generalizable social bot detection across heterogeneous datasets.}
  \label{fig:introduction-overview}
\end{figure}

\subsection{Challenges}

Deploying social bot detectors in practical environments presents two major challenges, as illustrated in Fig.~\ref{fig:introduction-overview}.

\textit{Challenge 1: How to construct consistent user representations from heterogeneous user evidence across datasets?}
Existing social bot detectors often rely on specific combinations of profile attributes, historical posts, behavioral statistics, and relational structures.
However, the organization and representation of such user evidence can vary substantially across datasets \cite{feng2021botrgcn, qiao2024dispelling}.
For example, different datasets may adopt different profile schemas, provide historical posts in different forms, or contain different types of relational information.
Such heterogeneity makes it difficult to directly reuse detectors that depend on fixed feature spaces or dataset-specific preprocessing.
Generalizable social bot detection therefore requires a representation mechanism that can organize heterogeneous user evidence into a consistent semantic form across datasets.

\textit{Challenge 2: How to preserve bot-human discriminability under distribution shifts across datasets?}
Even after heterogeneous user evidence is represented in a common form, models trained on source datasets may still capture dataset-specific regularities.
Differences in bot populations, topics, platform environments, and data collection procedures can induce distribution shifts across datasets.
Consequently, patterns that are predictive in source datasets may not remain reliable in unseen datasets.
Generalizable social bot detection therefore requires representations that reduce dependence on source-domain-specific information while preserving features that remain discriminative between bots and human users.

\subsection{Our Technical Overview and Contributions}

To conquer the above challenges, we propose \textbf{M}ulti-\textbf{G}ranularity Summarization and \textbf{D}omain-\textbf{I}nvariant \textbf{L}earning (MGDIL), a unified framework for generalizable social bot detection.
MGDIL follows a two-stage design.
To address \textit{Challenge 1}, it first maps heterogeneous user evidence into a shared semantic representation through multi-granularity summarization.
Specifically, profile attributes are converted into structured textual descriptions, while historical posts are summarized into compact user-level behavioral representations.
This design reduces dependence on dataset-specific feature schemas and provides a consistent semantic interface across datasets.
To address \textit{Challenge 2}, MGDIL further performs domain-invariant representation learning on the unified semantic representations.
Task-oriented instruction tuning captures discriminative semantics for social bot detection, domain-adversarial learning reduces dependence on source-domain-specific information, and cross-domain contrastive learning promotes class-consistent representations across different source domains.
Together, these objectives encourage representations that remain discriminative between bots and human users while being more transferable to unseen datasets.

The main contributions of this work are summarized as follows:

\begin{itemize}

\item
We propose MGDIL, a framework for generalizable social bot detection that jointly addresses representation inconsistency caused by heterogeneous user evidence and generalization degradation caused by cross-dataset distribution shifts.

\item We design a collaborative optimization scheme that integrates task-oriented LLM instruction tuning with domain-invariant representation learning. Specifically, instruction tuning enhances the model’s ability to capture fine-grained semantic cues and implicit camouflage patterns, while domain-adversarial training and contrastive learning
help mitigate cross-domain distribution shifts and promote stable, discriminative representations.

\item We construct a unified instruction-tuning
dataset from 15 social bot detection datasets through multigranularity summarization, train MGDIL on 13 datasets, and evaluate it on the remaining 2 datasets under distribution shift settings. Extensive experiments demonstrate that MGDIL achieves strong and consistent improvements in cross-domain generalization over competitive baselines.

\end{itemize}

\section{Related Work}

\subsection{Conventional Social Bot Detection Methods}

Conventional social bot detection methods can be broadly categorized into feature-based, behavioral and textual analysis-based, and graph-based approaches according to the user information they exploit.

\textbf{Feature-based methods.}
Early social bot detection studies mainly relied on feature engineering \cite{varol2018feature}, extracting statistical signals from profile metadata and basic posting statistics \cite{yang2020scalable}. These approaches typically used manually designed features together with traditional machine learning models to distinguish bots from humans \cite{singh2014big}. More recent studies have further incorporated neural architectures to improve representation learning from profile-level inputs. For instance, DeeProBot \cite{hayawi2022deeprobot} and TPBot\cite{yang2022new} adopted a deep neural network specifically designed for user profile features to detect malicious accounts.

\textbf{Behavioral and textual analysis-based methods.}
As social bots became increasingly adept at mimicking superficial profile characteristics \cite{cresci2020decade}, researchers shifted from profile-level features to richer behavioral modeling \cite{pozzana2020measuring,qiao2023social} and textual semantic analysis \cite{liu2023botmoe}. Cresci et al. \cite{cresci2016dna,cresci2017social} pioneered DNA-inspired behavioral modeling, which represents user actions as sequences to identify coordinated spambot groups.
Arin \cite{tifs2023} subsequently employ multiple LSTM components to capture complex behavioral patterns from user activity sequences.
Meanwhile, text-based neural methods employed architectures such as BiLSTMs \cite{wei2019twitter,kudugunta2018deep} and BERT \cite{guo2021social} to capture semantic inconsistencies and linguistic patterns in users' historical posts.

\textbf{Graph-based methods.}
More recently, graph neural network (GNN) based approaches have become a dominant paradigm for social bot detection by leveraging the relational structure of social networks \cite{de2021methods,dehghan2023detecting}. BotRGCN \cite{feng2021botrgcn} applies relational graph convolution to model heterogeneous user interactions, while RGT \cite{feng2022heterogeneity} further captures relation heterogeneity with graph transformers. Subsequent work has extended this direction by modeling dynamic interactions \cite{he2024dynamicity} and by jointly exploiting homophilic and heterophilic relations \cite{li2023multi,qiao2024dispelling}. Liu et al. \cite{tifs2025liu} further introduce a community-aware spatio-temporal hypergraph contrastive learning framework to capture higher-order interactions, community structures, and temporal dependencies among social accounts.

Most existing methods rely on specific combinations of user evidence, such as user profiles, historical posts, and social relations. However, these modalities are often incomplete, noisy, or unavailable in real-world settings, limiting the robustness of modality-dependent models \cite{de2021methods, ellaky2023systematic}.

\subsection{Generalizable Social Bot Detection Methods}

Although conventional social bot detection methods achieve promising performance when training and test data follow similar distributions, their performance often degrades when applied to previously unseen datasets or emerging bot populations \cite{ellaky2023systematic}.
Such generalization failures can arise from differences in topics, collection periods, platform environments, bot behaviors, feature schemas, and data acquisition procedures across datasets.
Consequently, recent studies have increasingly focused on improving the generalizability of social bot detectors beyond the datasets on which they are trained.

Existing efforts mainly improve generalization through adaptive learning, domain transfer, and multi-source knowledge integration.
For example, Song et al. \cite{song2025self} propose a self-supervised adaptive tuning method to mitigate performance degradation under distribution shifts by reducing representation discrepancies between source and target data.
Sun et al. \cite{sun2025adaptive} introduce AdaBot, which bridges the feature bias between source and target users through distribution alignment and target-aware adaptation.
Shi et al. \cite{shi2025bottrans} further propose BotTrans, a multi-source graph domain adaptation framework that exploits multiple source domains and selectively transfers task-relevant knowledge to a target network.

\begin{figure*}[t]
  \centering
  \includegraphics[width=\textwidth]{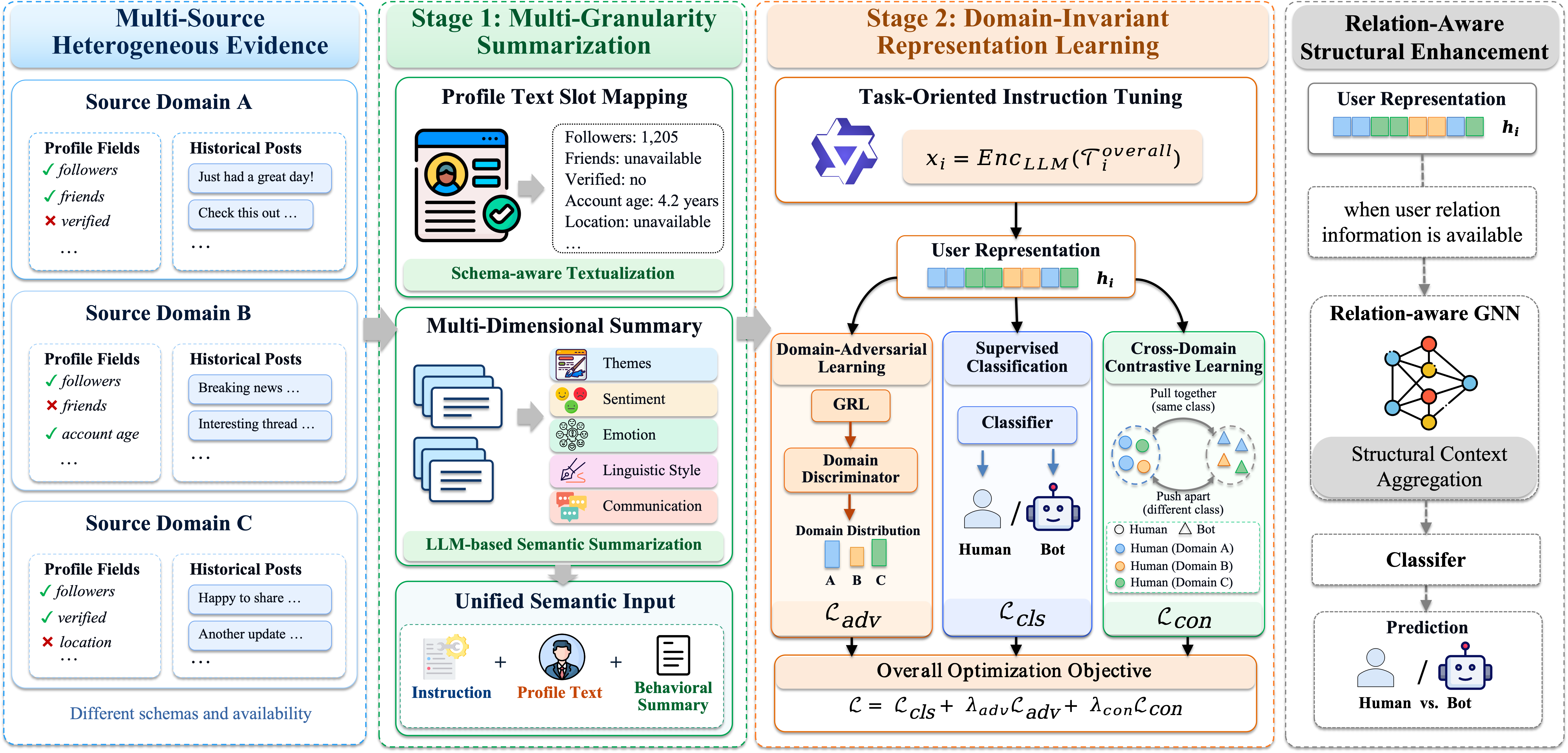}
  \caption{Overview of MGDIL. The framework first unifies heterogeneous profile and posting information through multi-granularity summarization, and then learns transferable user representations through domain-invariant representation learning. A relation-aware structural enhancement module is introduced when user relation information is available.}
  \label{fig:method_overview}
\end{figure*}

\section{Methodology}

We propose \textbf{MGDIL}, a cross-domain social bot detection framework that integrates multi-granularity summarization with domain-invariant representation learning, as illustrated in Fig.~\ref{fig:method_overview}. MGDIL follows a two-stage design. In the first stage, heterogeneous user evidence is transformed into a unified semantic representation by textualizing structured profile attributes and summarizing historical posts into compact behavioral descriptions. In the second stage, the unified semantic input is encoded by an instruction-tuned LLM and further optimized with supervised classification, domain-adversarial learning, and cross-domain contrastive learning. When user relation information is available, the learned user representations are subsequently processed by a relation-aware GNN to incorporate structural context before classification.

\subsection{Multi-Granularity Summarization}

User information collected from different social bot datasets may vary substantially in both schema and availability. For example, one dataset may contain profile attributes such as follower count, verification status, and account age, whereas another dataset may provide only a subset of these fields. Historical posts may also differ considerably in number, length, and content. Directly aligning such heterogeneous raw inputs therefore requires dataset-specific processing and assumes a consistent feature space across domains.

To reduce this dependence, MGDIL transforms heterogeneous user evidence into a unified textual representation at two complementary levels. At the profile level, structured user attributes are mapped into a slot-based textual format. At the behavior level, historical posts are summarized into a compact multi-dimensional semantic description. These representations are subsequently combined with a task instruction to construct a unified semantic input for domain-invariant representation learning.

\subsubsection{Profile Text Slot Mapping}
Let the profile schema contain \(J\) predefined fields,
\begin{equation}
\mathcal{F}=\{f_1,f_2,\ldots,f_J\}.
\end{equation}
For user \(i\in\{1,\ldots,N\}\), let \(v_{i,j}\) denote the observed value of field \(f_j\). Since field availability may vary across users and datasets, we define a binary indicator
\begin{equation}
m_{i,j}
=
\begin{cases}
1, & \text{if field } f_j \text{ is available for user } i,\\
0, & \text{otherwise}.
\end{cases}
\end{equation}

For an unavailable field, we assign a predefined placeholder
\(v^{(\mathrm{miss})}\), such as ``unavailable'' or ``unknown''. The normalized field value is defined as
\begin{equation}
\tilde{v}_{i,j}
=
\begin{cases}
v_{i,j}, & m_{i,j}=1,\\
v^{(\mathrm{miss})}, & m_{i,j}=0.
\end{cases}
\end{equation}

Each normalized field is subsequently converted into a natural-language slot through a field-specific rendering function \(\sigma_j(\cdot)\):
\begin{equation}
s_{i,j}
=
\sigma_j(\tilde{v}_{i,j}).
\end{equation}

The profile-level textual representation of user \(i\) is obtained by concatenating the rendered slots in a predefined order:
\begin{equation}
\mathcal{T}_{i}^{\mathrm{profile}}
=
\bigoplus_{j=1}^{J}
s_{i,j},
\end{equation}
where \(\bigoplus\) denotes string concatenation.

This formulation converts heterogeneous profile schemas into a consistent textual interface while explicitly retaining field-availability information. In particular, representing unavailable attributes with explicit placeholders avoids ambiguity between missing observations and observed field values. The complete profile-feature schema and rendering semantics are provided in the supplementary material.

\subsubsection{Multi-Dimensional Summary}
Historical posts provide rich semantic and behavioral evidence, but directly encoding all available posts introduces substantial variation in input length and content. We therefore summarize each user's posting history into a compact description that captures complementary characteristics of user behavior.
Let the historical post sequences of user \(i\) be denoted by \(P_i = (p_{i,1}, \dots, p_{i,T_i})\), where \(p_{i,t}\) is the text of the \(t\)-th post and \(T_i\) is the number of historical posts associated with user \(i\). Based on \(P_i\), we construct a multi-dimensional summary that captures the main semantic and expressive characteristics of the user's posting behavior.
We use LLMs (DeepSeek-V3.2 \cite{liu2025deepseek}) to summarize each user's historical posts along five dimensions: content themes (\textit{theme}) \cite{antypas2022twitter}, sentiment polarity (\textit{sent}) \cite{nakov2016semeval}, emotional tone (\textit{emo}) \cite{mohammad2018semeval}, linguistic style (\textit{style}) \cite{pavlick2016empirical,zampieri2019semeval}, and communicative function (\textit{func}) \cite{naaman2010really}. The full category definitions and prompt design are provided in the supplementary material. We denote the structured summary for user \(i\) as
\begin{equation}
    z_i^{\text{summ}} =
    \left(
    z_i^{\text{theme}},
    z_i^{\text{sent}},
    z_i^{\text{emo}},
    z_i^{\text{style}},
    z_i^{\text{func}}
    \right).
\end{equation}
For each dimension \(c \in \{\text{theme}, \text{sent}, \text{emo}, \text{style}, \text{func}\}\), the LLM summarizes \(P_i\) with respect to a predefined category space \(\mathcal{Y}^c\) and outputs the categories that best characterize the user's long-term posting behavior. The resulting fixed-format textual summary is written as:
\begin{equation}
    \mathcal{T}_{i}^{\text{summ}} = f_{\text{LLM}}\!\left(P_i; \{\mathcal{Y}^c\}_{c}\right),
\end{equation}
where \(f_{\text{LLM}}(\cdot)\) denotes the LLM-based summarization process under the category schema.

This transformation compresses variable-length posting histories into a compact and consistently formatted behavioral description, reducing direct dependence on the raw organization and length of historical posts. The complete category taxonomy, output constraints, and summarization prompts are provided in the supplementary material.

\subsubsection{Unified Semantic Input}

The profile representation and behavioral summary are combined with a task instruction to construct the unified semantic input:
\begin{equation}
\mathcal{T}_i^{\mathrm{overall}}
=
\mathcal{T}_i^{\mathrm{instruction}}
\oplus
\mathcal{T}_i^{\mathrm{profile}}
\oplus
\mathcal{T}_i^{\mathrm{summ}},
\end{equation}
where \(\mathcal{T}_i^{\mathrm{instruction}}\) specifies the social bot detection task.

Through this formulation, heterogeneous profile schemas and variable-length posting histories are mapped into a shared textual interface. The resulting representation serves as the input to the subsequent domain-invariant representation learning stage.

\subsection{Domain-Invariant Representation Learning}
Although multi-granularity summarization establishes a common semantic interface across datasets, the resulting representations may still encode source-domain-specific regularities associated with differences in data collection procedures, bot populations, topics, or dataset construction. Such correlations may be predictive within individual source domains but fail to generalize to unseen domains.

To mitigate this problem, MGDIL jointly optimizes the user representation with three complementary objectives. Supervised classification preserves information relevant to social bot detection, domain-adversarial learning reduces source-domain-predictive information, and cross-domain contrastive learning promotes class-consistent alignment across domains.

\subsubsection{Task-Oriented Instruction Tuning} We concatenate the instruction prompt \(\mathcal{T}_i^{\text{instruction}}\), the profile text \(\mathcal{T}_i^{\text{profile}}\), and the multi-dimensional summary of historical posts \(\mathcal{T}_i^{\text{summ}}\) into a unified instruction-formatted input:
\begin{equation}
    \mathcal{T}_i^{\text{overall}} =
    \mathcal{T}_i^{\text{instruction}}
    \oplus
    \mathcal{T}_i^{\text{profile}}
    \oplus
    \mathcal{T}_i^{\text{summ}}.
\end{equation}
We then employ Low-Rank Adaptation (LoRA) \cite{hu2022lora} to fine-tune the LLM under a joint optimization framework for domain-invariant learning. A complete instruction-formatted training instance is provided in the supplementary material.
After instruction fine-tuning, we use the LLM to encode the unified input \(\mathcal{T}_i^{\text{overall}}\) into a dense representation:
\begin{equation}
    \mathbf{x}_i = Enc_{\text{LLM}}\!\left(\mathcal{T}_i^{\text{overall}}\right),
\end{equation}
which is then used as the initial representation for subsequent domain-adversarial learning and cross-domain contrastive learning.

Each instruction-tuning instance consists of a task instruction, the rendered profile metadata, the multi-dimensional summary of historical posts, and the target account label. This format encourages the LLM to jointly reason over structured account attributes and summarized behavioral evidence. A complete instruction-tuning example is provided in the supplementary material.

\subsubsection{Domain-Adversarial Learning}
The latent representation \(\mathbf{h}_i\) may retain information that allows the corresponding source domain to be identified. We therefore introduce domain-adversarial learning to reduce dependence on such domain-predictive information.
Given the LLM-encoded representation \(\mathbf{x}_i\) of user \(i\), we first project it into a latent space:
\begin{equation}
    \mathbf{h}_i = f_{\theta}(\mathbf{x}_i) \in \mathbb{R}^{d},
\end{equation}
where \(f_{\theta}(\cdot)\) is an MLP parameterized by \(\theta\), and \(d\) is the hidden dimension. Based on \(\mathbf{h}_i\), we train a domain discriminator \(c_{\psi}(\cdot)\) to predict the domain label of each user. Assuming there are \(M\) domains, the predicted domain distribution is defined as
\begin{equation}
    \hat{\mathbf{p}}_i
    =
    \mathrm{softmax}\!\left(
        c_{\psi}\!\left(
            \mathrm{GRL}_{\lambda_{\mathrm{grl}}}(\mathbf{h}_i)
        \right)
    \right)
    \in \mathbb{R}^{M},
\end{equation}
where \(\psi\) denotes the parameters of the domain discriminator, \(\hat{\mathbf{p}}_i\) is the predicted probability distribution over the \(M\) domains, and \(d_i \in \{1,2,\dots,M\}\) is the ground-truth domain label of user \(i\).

To enable adversarial optimization, we insert a Gradient Reversal Layer (GRL) \cite{ganin2016domain} between the encoder and the domain discriminator. The GRL acts as the identity function in the forward pass, while reversing the gradient during backpropagation:
\begin{equation}
    \mathrm{GRL}_{\lambda_{\mathrm{grl}}}(x)=x,
\end{equation}
\begin{equation}
    \frac{\partial \,\mathrm{GRL}_{\lambda_{\mathrm{grl}}}(x)}{\partial x}
    =
    -\lambda_{\mathrm{grl}} I,
\end{equation}
where \(\lambda_{\mathrm{grl}} > 0\) is the gradient reversal coefficient and \(I\) is the identity matrix. This mechanism encourages the encoder to remove domain-specific signals while preserving task-relevant information.

The domain-adversarial loss is defined as
\begin{equation}
    \mathcal{L}_{\mathrm{adv}}
    =
    -\frac{1}{N}
    \sum_{i=1}^{N}
    \log p_{\psi}\!\left(
        d_i \mid \mathrm{GRL}_{\lambda_{\mathrm{grl}}}(\mathbf{h}_i)
    \right),
\end{equation}
where \(p_{\psi}(d_i \mid \mathrm{GRL}_{\lambda_{\mathrm{grl}}}(\mathbf{h}_i))\) denotes the probability assigned by the discriminator to the ground-truth domain label \(d_i\). Through adversarial optimization, the domain discriminator is trained to distinguish samples from different domains, while the encoder is encouraged to suppress domain-predictive information in the learned representations, thereby promoting more transferable representations across the source domains.

\subsubsection{Supervised Classification}

Reducing domain predictability alone does not guarantee that the resulting representation remains discriminative for social bot detection. We therefore simultaneously optimize \(\mathbf{h}_i\) using supervised account classification.

The classifier \(c_{\pi}(\cdot)\) predicts the probability distribution over the two account classes:
\begin{equation}
\hat{\mathbf{y}}_i
=
\mathrm{softmax}
\left(
c_{\pi}(\mathbf{h}_i)
\right)
\in
\mathbb{R}^{2},
\end{equation}
where \(\hat{\mathbf{y}}_i = [\hat{y}_{i,0}, \hat{y}_{i,1}]\) denotes the predicted probabilities for human and bot accounts, respectively.

The classification loss is defined as
\begin{equation}
\mathcal{L}_{\mathrm{cls}}
=
-\frac{1}{N}
\sum_{i=1}^{N}
\sum_{k\in\{0,1\}}
\mathbf{1}[y_i=k]
\log
\hat{y}_{i,k},
\end{equation}
where \(y_i\in\{0,1\}\) denotes the ground-truth account label.

This objective preserves discriminative information relevant to the social bot detection task while the other learning objectives regulate the structure of the representation space.

\subsubsection{Cross-Domain Contrastive Learning}
To further enhance class-level semantic alignment across domains, we introduce a cross-domain contrastive learning objective that encourages representations of users with the same class label to be closer across domains, while separating those with different class labels.
Specifically, we attach a projection head \(g_{\phi}(\cdot)\) to the latent representation $\mathbf{h}_i$ and normalize the projected output:
\begin{equation}
    \mathbf{u}_i
    =
    \frac{g_{\phi}(\mathbf{h}_i)}{\|g_{\phi}(\mathbf{h}_i)\|_2}
    \in \mathbb{R}^{r},
    \qquad
    \|\mathbf{u}_i\|_2 = 1,
\end{equation}
where \(g_{\phi}(\cdot)\) denotes the projection head parameterized by \(\phi\), and \(r\) is the dimension of the projected representation.

For each anchor user \(i\), we define a cross-domain positive set \(\mathcal{P}(i)\) and a negative set \(\mathcal{N}(i)\) as
\begin{equation}
    \mathcal{P}(i)
    =
    \left\{
    j \in \{1,\dots,N\}\setminus\{i\}
    \;:\;
    y_j = y_i,\;
    d_j \neq d_i
    \right\},
\end{equation}
\begin{equation}
    \mathcal{N}(i)
    =
    \left\{
    j \in \{1,\dots,N\}\setminus\{i\}
    \;:\;
    y_j \neq y_i
    \right\},
\end{equation}
where \(y_i\) is the class label of sample \(i\), and \(d_i\) is its domain label. Thus, \(\mathcal{P}(i)\) contains samples that share the same class label as the anchor but come from different domains, while \(\mathcal{N}(i)\) contains samples with different class labels.

Let $\mathcal{A}=\{i \in \{1,\dots,N\}: |\mathcal{P}(i)|>0\}$, denote the set of anchor samples that have at least one cross-domain positive sample. The cross-domain contrastive loss is defined as
\begin{equation}
    \mathcal{L}_{\mathrm{con}}
    =
    -\frac{1}{|\mathcal{A}|}
    \sum_{i \in \mathcal{A}}
    \frac{1}{|\mathcal{P}(i)|}
    \sum_{p \in \mathcal{P}(i)}
    \log
    \frac{
        \exp\!\left(\mathbf{u}_i^{\top}\mathbf{u}_p / \tau\right)
    }{
        \sum\limits_{a \in \mathcal{P}(i)\cup \mathcal{N}(i)}
        \exp\!\left(\mathbf{u}_i^{\top}\mathbf{u}_a / \tau\right)
    },
\end{equation}
where \(\tau > 0\) is a temperature hyperparameter, and \(\frac{1}{|\mathcal{P}(i)|}\) averages over all positive samples associated with anchor \(i\). Minimizing \(\mathcal{L}_{\mathrm{con}}\) encourages class-consistent alignment across domains in the learned representation space.

\subsubsection{Overall Optimization Objective}
To jointly optimize the three learning objectives, we define the overall loss as
\begin{equation}
    \mathcal{L}
    =
    \mathcal{L}_{\mathrm{cls}}
    +
    \lambda_{\mathrm{adv}} \mathcal{L}_{\mathrm{adv}}
    +
    \lambda_{\mathrm{con}} \mathcal{L}_{\mathrm{con}},
\end{equation}
where \(\mathcal{L}_{\mathrm{cls}}\), \(\mathcal{L}_{\mathrm{adv}}\), and \(\mathcal{L}_{\mathrm{con}}\) denote the classification, domain adversarial, and cross-domain contrastive losses, respectively. \(\lambda_{\mathrm{adv}}\) and \(\lambda_{\mathrm{con}}\) are hyperparameters that balance the contributions of the latter two objectives. This joint optimization encourages the learned representations to be both domain-invariant and class-discriminative.

\subsection{Relation-Aware Structural Enhancement}
When user relation information is available, we further extend MGDIL with a relation-aware graph module to incorporate structural context.

Specifically, we construct a user relation graph
\begin{equation}
    \mathcal{G} = (\mathcal{V}, \mathcal{E}, \mathcal{R}),
\end{equation}
where $\mathcal{V}$ is the set of user nodes, $\mathcal{E}$ is the set of edges, and $\mathcal{R}$ is the set of relation types. For each user node $v \in \mathcal{V}$, we initialize its node feature using the representation learned by the previous module:
\begin{equation}
    \hat{\mathbf{h}}_v^{(0)} = \mathbf{h}_v,
\end{equation}
where $\mathbf{h}_v \in \mathbb{R}^{d}$ denotes the semantic representation of user $v$ obtained from the preceding representation learning module.

To incorporate structural dependencies among users, we apply a message-passing mechanism to refine node representations. At the $l$-th layer, node $v$ aggregates information from its neighbors under different relation types:
\begin{equation}
    \mathbf{m}_v^{(l)}
    =
    \sum_{r \in \mathcal{R}}
    \mathrm{AGG}_{r}^{(l)}
    \left(
    \left\{
    \mathbf{W}_{r}^{(l)} \hat{\mathbf{h}}_u^{(l)}
    \; \middle| \;
    u \in \mathcal{N}_r(v)
    \right\}
    \right),
\end{equation}
where $\mathcal{N}_r(v)$ denotes the set of neighbors of node $v$ under relation type $r$, $\mathbf{W}_{r}^{(l)}$ is a relation-specific transformation matrix, and $\mathrm{AGG}_{r}^{(l)}(\cdot)$ is a relation-aware aggregation function, such as mean pooling, sum pooling, or attention-based aggregation. The node representation is then updated by combining the current node representation with the aggregated neighborhood message:
\begin{equation}
    \hat{\mathbf{h}}_v^{(l+1)}
    =
    \mathrm{UPD}^{(l)}
    \left(
        \hat{\mathbf{h}}_v^{(l)}, \mathbf{m}_v^{(l)}
    \right),
\end{equation}
where $\mathrm{UPD}^{(l)}(\cdot)$ denotes the node update function. After $L$ layers of propagation, the final graph-enhanced representation of user node $v$ is given by $\mathbf{z}_v = \hat{\mathbf{h}}_v^{(L)}$.

The graph-enhanced representation $\mathbf{z}_v$ is then used for downstream account classification. Specifically, the predicted class distribution for user node $v$ is computed as
\begin{equation}
    \hat{\mathbf{y}}_v = \mathrm{softmax}(g_{\pi}(\mathbf{z}_v)) \in \mathbb{R}^{2},
\end{equation}
where \(g_{\pi}(\cdot)\) denotes the classifier parameterized by \(\pi\), and \(\hat{\mathbf{y}}_v = [\hat{y}_{i,0}, \hat{y}_{i,1}]\) is the predicted probability vector for node \(v\). The graph-based classifier is optimized with the cross-entropy loss:
\begin{equation}
    \mathcal{L}_{\mathrm{gcls}}
    =
    - \sum_{v=1}^{N} \sum_{k \in \{0,1\}}
    \mathbf{1}[y_v = k] \log \hat{y}_{v,k},
\end{equation}
where $N$ is the number of user nodes, $y_v$ is the ground-truth label of node $v$, and $\hat{y}_{v,k}$ is the probability that node $v$ belongs to class $k$.

\begin{table*}[t]
\centering
\small
\setlength{\tabcolsep}{5pt}
\caption{Main results on the merged dataset, which is constructed by combining TwiBot-2020 and Fox-2023, as well as on the individual TwiBot-2020 and Fox-2023 datasets. ``MetaData'' uses structured user profile metadata as input, while ``Meta-Summary'' uses both profile metadata and summarized user-history posts. Since only TwiBot-2020 provides user relation information for graph construction, GNN-based baselines are evaluated only on TwiBot-2020.}
\label{tab:main_comparison}
\resizebox{\textwidth}{!}{
\begin{tabular}{ll|cc|cc|cc}
\toprule
\multicolumn{2}{c}{\multirow{2}{*}{Method}}
& \multicolumn{2}{c}{Merged Dataset}
& \multicolumn{2}{c}{Twibot-2020}
& \multicolumn{2}{c}{Fox-2023} \\
\cmidrule(lr){3-4} \cmidrule(lr){5-6} \cmidrule(lr){7-8}
\multicolumn{2}{c}{}
& Accuracy & Macro F1
& Accuracy & Macro F1
& Accuracy & Macro F1 \\
\midrule

\multirow{4}{*}{\textbf{ML Methods}}
& Random Forest & 56.43 $\pm$ 1.20 & 56.35 $\pm$ 1.25 & 60.42 $\pm$ 1.43 & 60.37 $\pm$ 1.45 & 24.85 $\pm$ 1.34 & 23.43 $\pm$ 1.48 \\
& SVM           & 54.37 $\pm$ 0.30 & 54.04 $\pm$ 0.28 & 57.29 $\pm$ 0.26 & 56.96 $\pm$ 0.28 & 31.24 $\pm$ 0.60 & 30.92 $\pm$ 0.55 \\
& Decision Tree & 47.89 $\pm$ 0.39 & 46.45 $\pm$ 0.50 & 49.43 $\pm$ 0.52 & 47.69 $\pm$ 0.69 & 35.69 $\pm$ 0.88 & 35.66 $\pm$ 0.89 \\
& AdaBoost      & 65.93 $\pm$ 0.17 & 64.04 $\pm$ 0.16 & 70.78 $\pm$ 0.13 & 68.47 $\pm$ 0.11 & 27.49 $\pm$ 0.53 & 26.73 $\pm$ 0.59 \\
\midrule

\multirow{3}{*}{\textbf{NN Methods}}
& MLP          & 69.41 $\pm$ 4.37 & 68.42 $\pm$ 3.80 & 71.29 $\pm$ 3.35 & 70.34 $\pm$ 2.85 & 54.47 $\pm$ 23.41 & 52.36 $\pm$ 21.85 \\
& RoBERTa & 56.79 $\pm$ 1.33 & 51.90 $\pm$ 0.27 & 52.69 $\pm$ 1.51 & 48.87 $\pm$ 0.34 & 89.35 $\pm$ 0.48 & 79.86 $\pm$ 0.99 \\
& MLP + RoBERTa & 69.58 $\pm$ 4.51 & 60.96 $\pm$ 9.49 & 67.98 $\pm$ 5.06 & 60.66 $\pm$ 9.85 & 82.21 $\pm$ 1.68 & 53.34 $\pm$ 7.33 \\
\midrule

\multirow{3}{*}{\textbf{GNN Methods}}
& BotRGCN
& 62.70 $\pm$ 6.97 & 55.33 $\pm$ 11.37
& 62.70 $\pm$ 6.97 & 55.33 $\pm$ 11.37
& -- & -- \\

& RGT
& 69.87 $\pm$ 9.39 & 66.93 $\pm$ 11.74
& 69.87 $\pm$ 9.39 & 66.93 $\pm$ 11.74
& -- & -- \\

& HGT
& 63.35 $\pm$ 13.23 & 55.69 $\pm$ 18.15
& 63.35 $\pm$ 13.23 & 55.69 $\pm$ 18.15
& -- & -- \\
\midrule

\multirow[c]{3}{*}{
    \makecell[c]{
        \textbf{Cross-domain}\\
        \textbf{Methods}
    }
}
& BotRGCN + Tuning       & 63.28 $\pm$ 3.01                & 62.49 $\pm$ 8.89                & 63.28 $\pm$ 3.01 & 62.49 $\pm$ 8.89 & --                & --                \\
& RGT + Tuning          & 67.68 $\pm$ 2.85                & 66.60 $\pm$ 2.37                & 67.68 $\pm$ 2.85 & 66.60 $\pm$ 2.37 & --                & --                \\
& AdaBot        & 64.40 $\pm$ 0.90                & 74.20 $\pm$ 0.80                & 64.40 $\pm$ 0.90 & 74.20 $\pm$ 0.80 & --                & --                \\
\midrule

\multirow{3}{*}{\textbf{LLM Methods}}
& GPT-5.4-nano       & 46.60 $\pm$ 0.05 & 46.34 $\pm$ 0.05 & 47.51$\pm$ 0.07  & 47.13 $\pm$ 0.09 & 39.53 $\pm$ 0.37 & 39.38 $\pm$ 0.38     \\
& Deepseek-V3.2          & 60.23 $\pm$ 0.25 & 40.24 $\pm$ 0.14 & 57.42$\pm$ 0.26  & 38.52 $\pm$ 0.13 & 80.01 $\pm$ 0.51 & 56.87 $\pm$ 1.29     \\
& Qwen2.5-1.5B-Instruct          & 54.29 $\pm$ 0.00 & 53.07 $\pm$ 0.00 & 54.69$\pm$ 0.00  & 53.49 $\pm$ 0.00 & 51.16 $\pm$ 0.00 & 49.86 $\pm$ 0.00     \\
\midrule

\multirow{2}{*}{\textbf{LoRA-Finetune}}
& MetaData      & 72.07 $\pm$ 3.91 & 69.50 $\pm$ 4.11 & 76.64 $\pm$ 2.72 & 73.74 $\pm$ 3.25 & 36.16 $\pm$ 15.85 & 34.86 $\pm$ 16.19 \\
& Meta-Summary  & 77.08 $\pm$ 1.24 & 73.42 $\pm$ 1.53 & 77.37 $\pm$ 1.40 & 73.02 $\pm$ 1.96 & 75.86 $\pm$ 20.22 & 70.41 $\pm$ 17.18 \\
\midrule

\multirow{2}{*}{\textbf{MGDIL (Ours)}}
& MetaData      & 81.01 $\pm$ 0.28 & 77.89 $\pm$ 0.26 & 79.80 $\pm$ 0.10 & \textbf{76.82 $\pm$ 0.10} & 90.45 $\pm$ 2.89 & 87.25 $\pm$ 3.34 \\
& Meta-Summary  & \textbf{81.63 $\pm$ 0.11} & \textbf{78.27 $\pm$ 0.21} & \textbf{79.83 $\pm$ 0.18} & 76.71 $\pm$ 0.26 & \textbf{96.42 $\pm$ 0.60} & \textbf{94.13 $\pm$ 0.84} \\
\bottomrule
\end{tabular}
}
\end{table*}

\section{Experiments}

\subsection{Experimental Details}
\subsubsection{Datasets}
To ensure broad coverage of social bot behaviors and data conditions, we construct a unified instruction-tuning dataset from 15 social bot detection datasets.
Specifically, we collect 14 datasets from the Bot Repository\footnote{\url{https://botometer.osome.iu.edu/bot-repository/datasets.html}}, including Cresci-2015 \cite{cresci2015fame}, Cresci-2017 \cite{cresci2017social}, Gilani-2017 \cite{gilani2017bots}, Cresci-stock-2018 \cite{cresci2018fake}, Midterm-2018 \cite{yang2020scalable}, Botometer-Feedback-2019 \cite{yang2019arming}, Botwiki-2019 \cite{yang2020scalable}, Celebrity-2019 \cite{yang2019arming}, Cresci-RTbust-2019 \cite{mazza2019rtbust}, Political-bots-2019 \cite{yang2019arming}, Pornbots-2019 \cite{yang2019arming}, Vendor-purchased-2019 \cite{yang2019arming}, Verified-2019 \cite{yang2020scalable}, and Twibot-2020 \cite{feng2021twibot}. We further include Fox-2023 \cite{yang2023anatomy}, a recent dataset of LLM-driven malicious social bots, to evaluate robustness under newer and shifted distributions. Detailed dataset-level modality coverage, annotation protocols, and field availability are provided in the supplementary material.

\textbf{Implementation Details.}
We adopt Qwen2.5-1.5B \cite{hui2024qwen2} as the backbone LLM and fine-tune it with LoRA \cite{hu2022lora}. Experiments are conducted on two NVIDIA H800 GPUs with 80\,GB memory each under CUDA 12.8. To ensure the reliability of the experimental results, we repeat each experiment using five different random seeds. Model performance is evaluated in terms of Accuracy and Macro-F1. Further implementation details and hyperparameter settings are provided in the supplementary material.

\textbf{Baselines.}
\label{sec:baselines}
We compare MGDIL with representative baselines from six methodological families. Since different methods require different input modalities and structural information, we follow the feasible input setting of each method and explicitly distinguish modality-specific comparisons from controlled comparisons using the same input representation.

\begin{itemize}[leftmargin=*,labelsep=0.5em,itemsep=2pt,topsep=3pt]

    \item \textbf{Traditional machine learning methods (ML).}
    Random Forest, Support Vector Machine (SVM), Decision Tree, and AdaBoost are included following metadata-based social bot detection settings \cite{yang2020scalable}.

    \item \textbf{Neural network methods (NN).}
    We include MLP, RoBERTa, and MLP+RoBERTa \cite{liu2019roberta,qiao2025botsim}. MLP uses metadata features, RoBERTa uses historical posts, and MLP+RoBERTa combines both features. RoBERTa and MLP+RoBERTa are trained only on datasets for which historical post content is available.

    \item \textbf{Graph neural network methods (GNN).}
    We include BotRGCN \cite{feng2021botrgcn}, RGT \cite{feng2022heterogeneity}, and HGT \cite{hu2020heterogeneous}. Since these methods require user-relation graphs, they are evaluated only when compatible relational information is available.

    \item \textbf{Cross-domain methods.}
    We include Tuning \cite{song2025self} and AdaBot \cite{sun2025adaptive} as representative methods for social bot detection under distribution shifts, evaluated under their applicable configurations.

    \item \textbf{Prompt-based LLM methods.}
    GPT-5.4-nano, DeepSeek-V3.2 \cite{liu2025deepseek}, and Qwen2.5-1.5B-Instruct \cite{qwen2024qwen25} are evaluated using direct prompting without task-specific parameter adaptation.

    \item \textbf{Instruction-tuning baseline.}
    We construct a LoRA-Finetune baseline \cite{hu2022lora} using the same backbone, instruction format, and training data as MGDIL, while removing domain-adversarial and cross-domain contrastive learning. This provides a direct controlled comparison for evaluating the proposed domain-invariant learning strategy.

\end{itemize}

\subsection{Comparison with State-of-the-Art Methods}
We evaluate MGDIL on two target datasets under the cross-domain setting and compare it with the representative baselines introduced in Section~\ref{sec:baselines}. Table~\ref{tab:main_comparison} summarizes the overall results. We draw the following observations:

(1) \textbf{Cross-domain social bot detection remains challenging under heterogeneous evaluation conditions.}
Traditional ML methods perform poorly across most evaluation settings, especially on Fox-2023, where both Accuracy and Macro-F1 are extremely low. NN-based methods show moderate improvements over ML baselines in some cases, but remain unstable, as indicated by the large standard deviations of MLP and MLP+RoBERTa. GNN-based and Cross-domain methods achieve relatively better results on TwiBot-2020, and Cross-domain adaptation further improves performance; however, these methods rely on graph availability or specific adaptation settings, so their coverage is narrower than that of metadata- and summary-based approaches. Overall, the results suggest that existing baselines remain sensitive to modality availability and distribution shift.

(2) \textbf{Off-the-shelf LLMs and vanilla instruction tuning provide stronger semantic modeling, but are still insufficient for robust cross-domain generalization.} Directly prompted LLMs, including GPT-5.4-nano, Deepseek-V3.2 and Qwen2.5-1.5B-Instruct, perform poorly on all target datasets, suggesting that off-the-shelf LLMs alone cannot effectively solve this task. In contrast, LoRA-Finetune consistently outperforms most conventional baselines on the merged dataset and TwiBot-2020, demonstrating the benefit of task-specific instruction tuning. However, vanilla instruction tuning remains unstable under more severe distribution shifts. In particular, LoRA-Finetune (MetaData) performs poorly on Fox-2023, while LoRA-Finetune (Meta-Summary) shows very large standard deviations, indicating limited robustness without explicit domain-invariant learning.

(3) \textbf{MGDIL provides strong performance under the evaluated cross-domain settings.} MGDIL achieves the strongest or near-strongest results among methods that can be evaluated under the corresponding modality conditions. On the merged dataset, it outperforms the best LoRA-Finetune baseline in both Accuracy and Macro-F1. On Twibot-2020, both MGDIL variants achieve competitive results, with Meta-Summary performing best in Accuracy and MetaData in Macro-F1. The most substantial improvement is observed on Fox-2023, where MGDIL markedly surpasses the available baselines. Moreover, MGDIL exhibits smaller standard deviations than competing methods in several settings, suggesting more stable behavior under the considered distribution shifts.

(4) \textbf{Summary-based user representations are particularly beneficial under severe domain shifts.}
The Meta-Summary variant generally outperforms the MetaData variant, especially on the merged dataset and Fox-2023, suggesting that multi-granularity summaries provide useful user-level behavioral signals beyond metadata alone. This advantage becomes particularly evident when the target domain differs substantially from the source domains, as in Fox-2023. At the same time, MetaData remains competitive on TwiBot-2020 and even achieves the best Macro-F1, suggesting that structured metadata still provides useful complementary information. Overall, summary-based representations appear beneficial under stronger shifts, while metadata can still be effective in specific target domains.

\begin{figure}[t]
  \centering
  \includegraphics[width=\columnwidth]{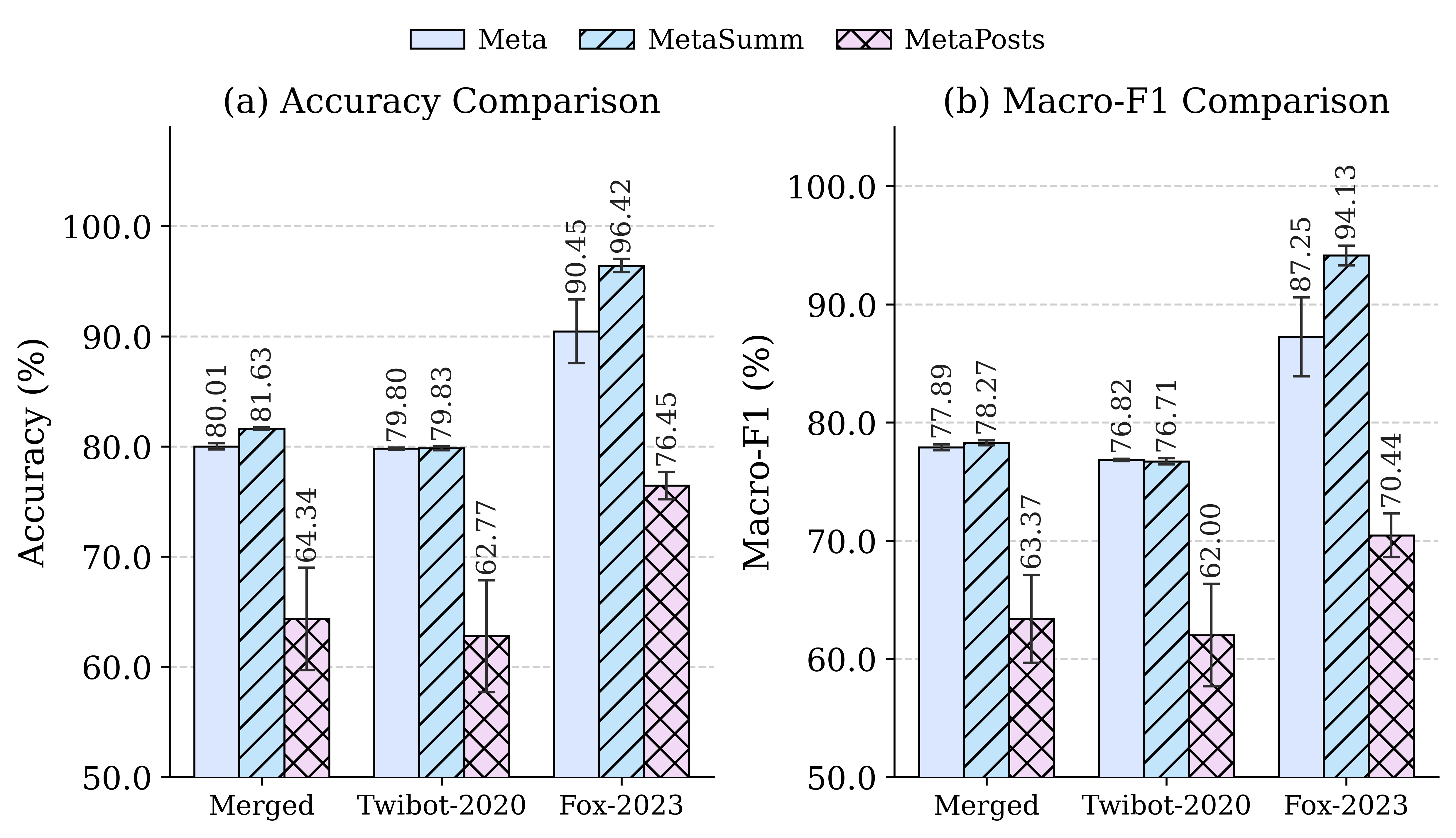}
  \caption{Comparison of Raw Historical Posts and Summarized Historical Posts.}
  \label{fig:recent_summary}
\end{figure}

\begin{table*}[t]
\centering
\caption{Ablation study of the domain-adversarial learning and contrastive learning modules on the merged dataset, TwiBot-2020, and Fox-2023, with individual and joint removal results.}
\label{tab:ablation_transposed}
\setlength{\tabcolsep}{8pt}
\renewcommand{\arraystretch}{1.1}
\begin{tabular}{llcccc}
\toprule
\textbf{Dataset} & \textbf{Metric}& \textbf{W/O Adv. \& Con.} & \textbf{W/O Adversarial} & \textbf{W/O Contrast} & \textbf{MGDIL} \\
\midrule
\multirow{2}{*}{Merged}
& Accuracy (\%) & 78.91 $\pm$ 1.82 & 80.24 $\pm$ 1.06 & 81.29 $\pm$ 0.33 & \textbf{81.63 $\pm$ 0.11} \\
& Macro-F1 (\%) & 75.85 $\pm$ 1.69 & 76.18 $\pm$ 0.84 & 78.14 $\pm$ 0.22 & \textbf{78.27 $\pm$ 0.21} \\
\midrule
\multirow{2}{*}{Twibot-2020}
& Accuracy (\%) & 77.07 $\pm$ 2.05 & 78.93 $\pm$ 1.11 & 79.27 $\pm$ 0.53 & \textbf{79.83 $\pm$ 0.18} \\
& Macro-F1 (\%) & 74.42 $\pm$ 1.89 & 75.74 $\pm$ 1.45 & 76.39 $\pm$ 0.44 & \textbf{76.71 $\pm$ 0.26} \\
\midrule
\multirow{2}{*}{Fox-2023}
& Accuracy (\%) & 93.72 $\pm$ 1.61 & 95.08 $\pm$ 1.51 & 95.88 $\pm$ 0.83 & \textbf{96.42 $\pm$ 0.60} \\
& Macro-F1 (\%) & 89.37 $\pm$ 3.01 & 91.52 $\pm$ 2.84 & 93.60 $\pm$ 2.32 & \textbf{94.13 $\pm$ 0.84} \\
\bottomrule
\end{tabular}
\end{table*}

\subsection{Ablation Study}
\label{sec:ablation}
\subsubsection{Impact of Multi-Granularity Summarization}

To evaluate the effectiveness of multi-granularity summarization of historical posts, we conduct an ablation study under three input settings, as shown in Figure~\ref{fig:recent_summary}: (1) metadata only (\textbf{Meta}), (2) metadata combined with summarized historical posts (\textbf{MetaSumm}), and (3) metadata combined with raw historical posts (\textbf{MetaPosts}). The results show that directly using raw historical posts leads to worse performance than using summarized historical information. In some cases, incorporating raw posts even weakens the benefit of metadata features. A likely reason is that raw posting histories contain substantial noise, redundancy, and irrelevant content, which may obscure informative user-level patterns and make optimization more difficult. In contrast, multi-granularity summarization compresses long posting histories into a more compact and informative representation, enabling the model to capture the most relevant signals for bot detection more effectively. Overall, these findings confirm that the summarization step plays an important role in improving detection performance.

To better understand why summarized historical posts improve performance, we further examine four datasets with historical posting sequences, namely Cresci-2015, Cresci-2017, TwiBot-2020, and Fox-2023, along five dimensions: emotional tone, sentiment polarity, linguistic style, content theme, and communicative function. The analysis shows that recent bot datasets exhibit stronger shifts toward emotionally intense expression, more negative or mixed sentiment, more aggressive and formalized language, issue-centered topics such as cryptocurrency and politics, and more broadcast-oriented communicative functions. These observations suggest that raw posts contain substantial temporal and topical variation, while multi-granularity summarization can compress such variation into more compact user-level behavioral signals. Detailed heatmaps, grouped-bar visualizations, and per-dimension analyses are provided in the supplementary material.

\subsubsection{Impact of Domain-Invariant Representation Learning}
To evaluate the contribution of domain-adversarial learning and cross-domain contrastive learning, we conduct ablation studies by removing each module individually (`W/O Adversarial' and `W/O Contrast') and jointly (`W/O Adv. \& Con.') while keeping all other settings unchanged. Table~\ref{tab:ablation_transposed} shows that removing either module consistently degrades performance across all evaluation sets, indicating that both components contribute to MGDIL. In particular, removing domain-adversarial learning results in a larger degradation than removing contrastive learning, suggesting that it plays an important role in reducing domain-predictive variation. Moreover, removing both modules leads to a further performance drop, indicating that the two objectives provide complementary signals. Specifically, domain-adversarial learning discourages domain-discriminative information, while cross-domain contrastive learning improves class-level alignment across domains.

\begin{table}[t]
\centering
\caption{Performance comparison between the original GNN-based models and their MGDIL-enhanced variants, trained on Cresci-2015 and evaluated on TwiBot-2020.}
\label{tab:gnn_comparison}
\setlength{\tabcolsep}{4pt}
\renewcommand{\arraystretch}{1.12}
\begin{tabular}{llccc}
\toprule
\textbf{Model} & \textbf{Metric} & \textbf{Original}
& \textbf{Enhanced} & \textbf{Gain (pp)} \\
\midrule

\multirow{2}{*}{GCN}
& Accuracy (\%)  & 56.07 $\pm$ 15.54
& \textbf{79.13 $\pm$ 0.43} & $\uparrow$23.06 \\
& Macro-F1 (\%) & 43.92 $\pm$ 19.09
& \textbf{76.29 $\pm$ 0.39} & $\uparrow$32.37 \\

\multirow{2}{*}{GAT}
& Accuracy (\%)  & 63.99 $\pm$ 16.98
& \textbf{79.18 $\pm$ 0.65} & $\uparrow$15.19 \\
& Macro-F1 (\%) & 57.82 $\pm$ 18.90
& \textbf{76.29 $\pm$ 0.50} & $\uparrow$18.47 \\

\midrule

\multirow{2}{*}{BotRGCN}
& Accuracy (\%)  & 62.70 $\pm$ 6.97
& \textbf{79.46 $\pm$ 0.32} & $\uparrow$16.76 \\
& Macro-F1 (\%) & 55.33 $\pm$ 11.37
& \textbf{76.60 $\pm$ 0.27} & $\uparrow$21.27 \\

\multirow{2}{*}{RGT}
& Accuracy (\%)  & 69.87 $\pm$ 9.39
& \textbf{79.64 $\pm$ 0.23} & $\uparrow$9.77 \\
& Macro-F1 (\%) & 66.93 $\pm$ 11.74
& \textbf{76.69 $\pm$ 0.25} & $\uparrow$9.76 \\

\multirow{2}{*}{HGT}
& Accuracy (\%)  & 63.35 $\pm$ 13.23
& \textbf{79.54 $\pm$ 0.09} & $\uparrow$16.19 \\
& Macro-F1 (\%) & 55.69 $\pm$ 18.15
& \textbf{76.65 $\pm$ 0.19} & $\uparrow$20.96 \\

\bottomrule
\end{tabular}
\end{table}

\begin{figure*}[t]
\centering

\subfloat[Human-bot differences in communicative function.]
{
\includegraphics[width=0.78\linewidth]
{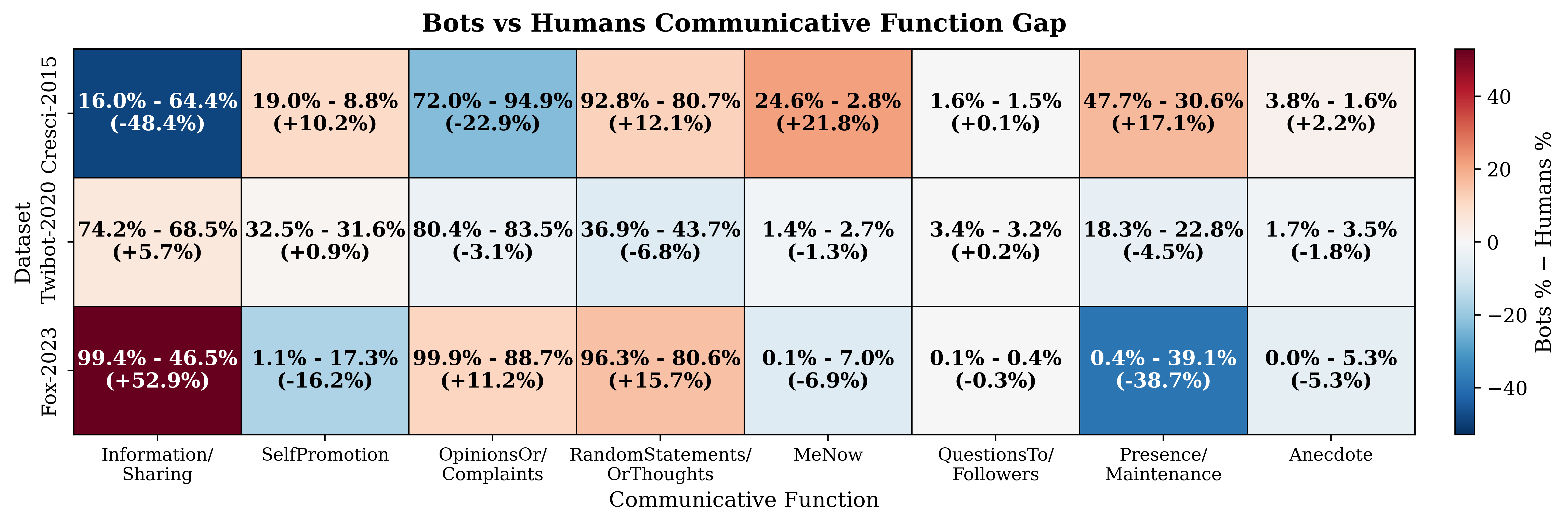}
}
\hfill
\subfloat[Communicative function distribution of bots across datasets.]
{
\includegraphics[width=0.78\linewidth]
{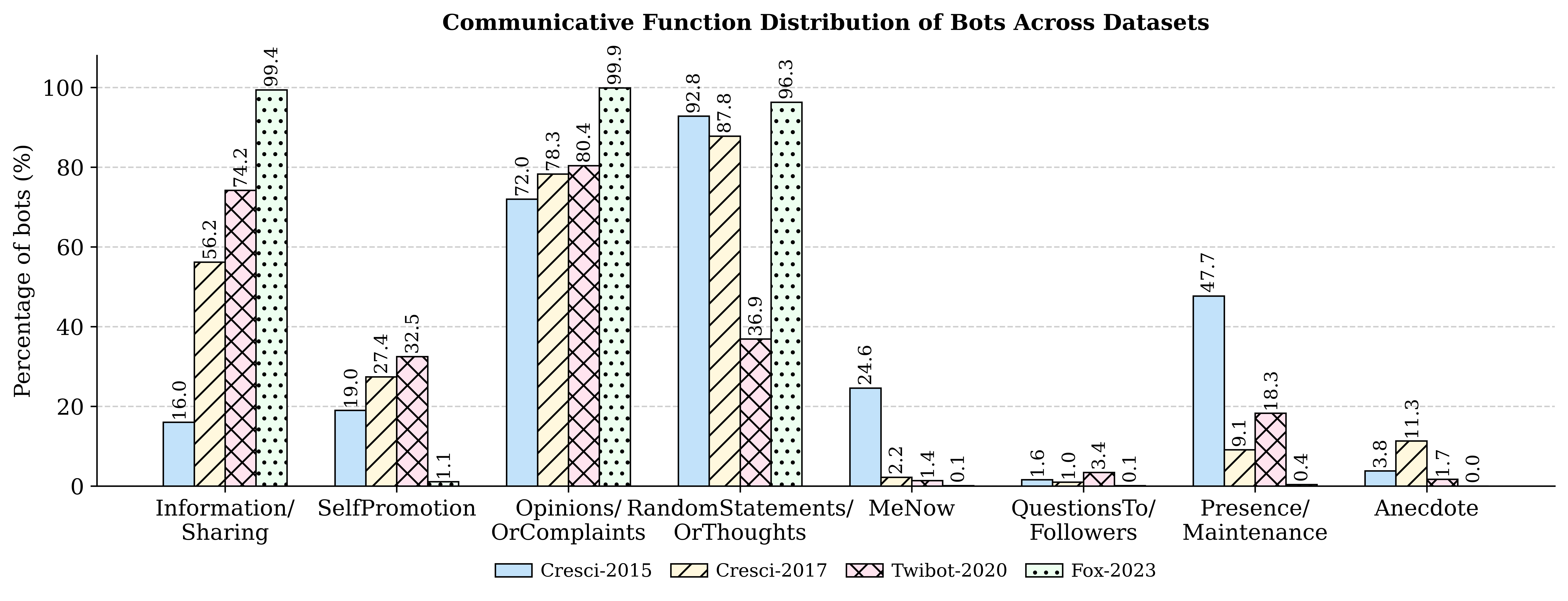}
}

\caption{
Comparison of communicative functions between human and bot accounts across datasets.
}
\label{fig:function_analysis}

\end{figure*}
\subsection{Transferability to Graph-based Detectors}
To avoid making graph learning appear as an additional core mechanism, we treat the GNN experiment as an embedding-transfer study. We integrate the fine-tuned embeddings learned by MGDIL into five representative GNN models (Enhanced): GCN and GAT for homogeneous graphs, and BotRGCN \cite{feng2021botrgcn}, RGT \cite{feng2022heterogeneity}, and HGT for heterogeneous graphs, and compare them with their original versions (Original). For the original GNN baselines, user representations are derived from metadata features encoded by an MLP and historical post features encoded by RoBERTa with mean pooling. For the MGDIL-enhanced variants, we replace these input features with the fine-tuned embeddings $\mathbf{h}_i$ of user $i$ produced by MGDIL while keeping the GNN architectures unchanged. As shown in Table~\ref{tab:gnn_comparison}, all enhanced GNNs improve substantially over their original versions, suggesting that MGDIL embeddings can serve as transferable input features for both homogeneous and heterogeneous graph-based detectors when graph structure is available. Although the use of MGDIL embeddings leads to substantial performance gains in both settings, the homogeneous GNNs still achieve slightly lower performance than the heterogeneous GNNs, indicating that explicitly modeling heterogeneous relational information can provide additional benefits. This study therefore supports the compatibility of the learned representations with downstream relational models, rather than introducing a separate graph-learning contribution.

\begin{figure*}[t]
\centering

\subfloat[Human-bot differences in emotional tone.]
{
\includegraphics[width=0.48\linewidth]
{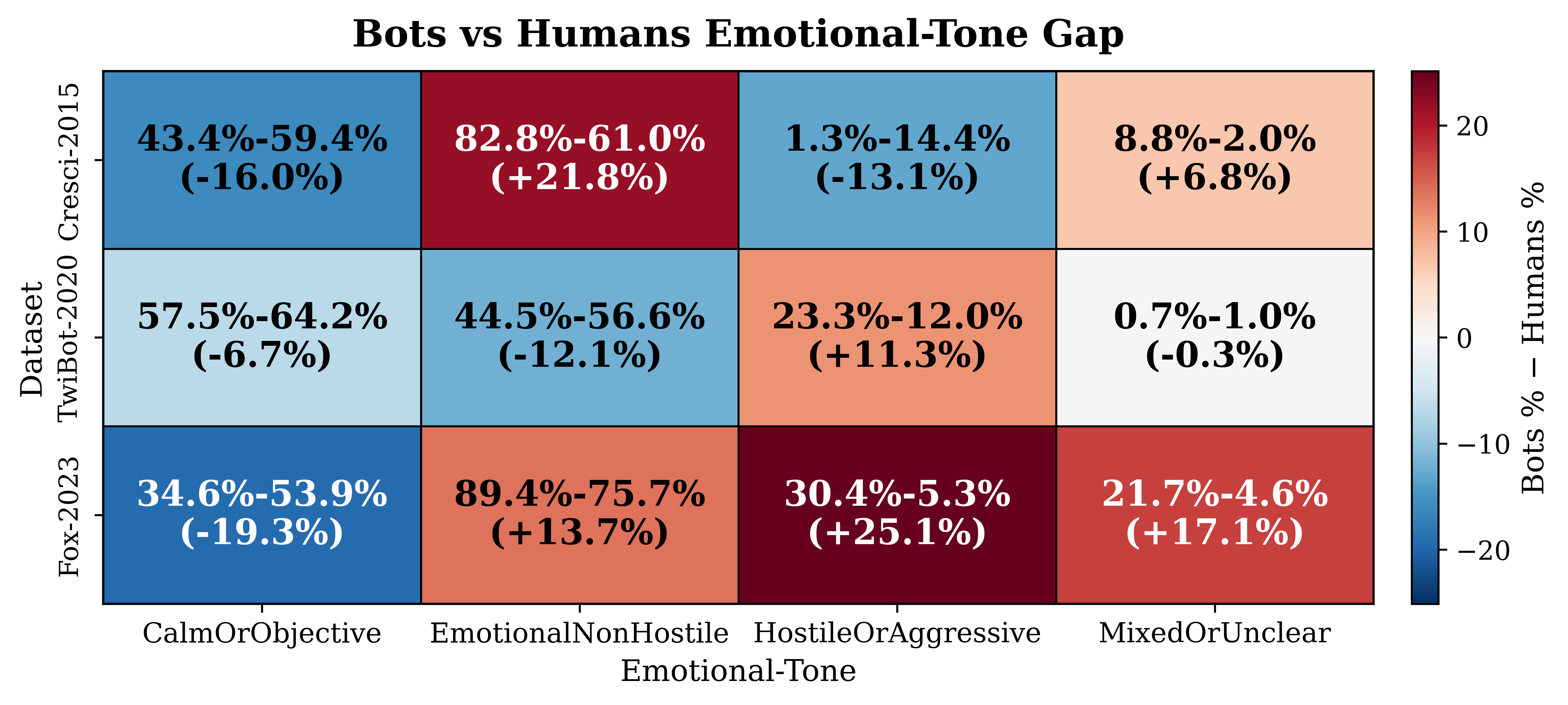}
}
\hfill
\subfloat[Emotional tone distribution of bots across datasets.]
{
\includegraphics[width=0.48\linewidth]
{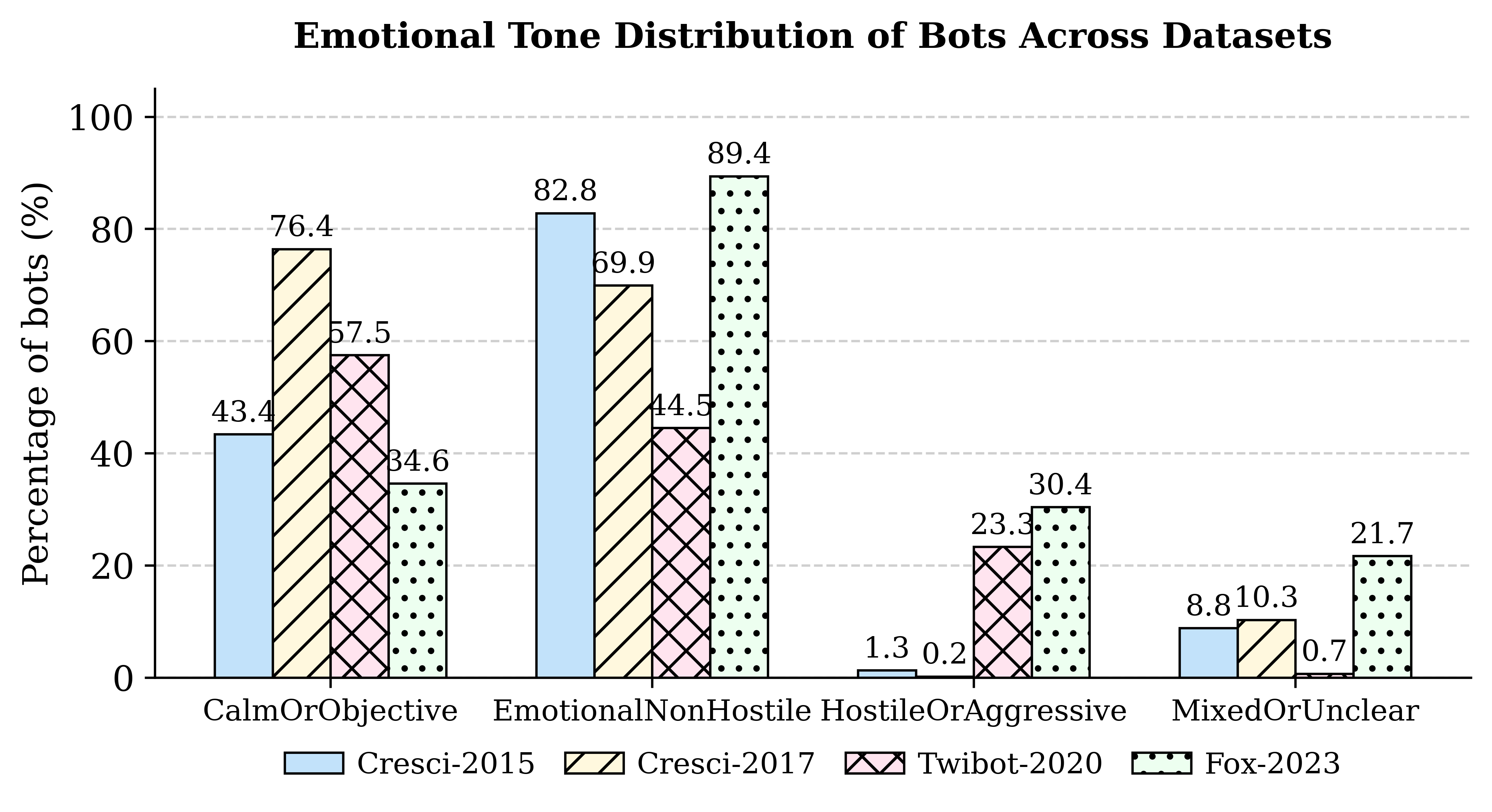}
}

\caption{
Comparison of emotional tone between humans and bots across datasets.
}
\label{fig:emotion_analysis}

\end{figure*}

\begin{table*}[t]
\centering
\caption{Robustness to additional random masking of originally available
profile features. The Full setting preserves the native missing values,
while other masking rates are applied on top of the original incompleteness.
Results are reported as mean $\pm$ standard deviation over five runs (\%).}
\label{tab:missing_feature_robustness}
\setlength{\tabcolsep}{6.5pt}
\renewcommand{\arraystretch}{1.12}

\begin{tabular}{c|cc|cc|cc}
\toprule
\multirow{2}{*}{\makecell{Additional\\Masking Rate}}
& \multicolumn{2}{c|}{\textbf{Merged}}
& \multicolumn{2}{c|}{\textbf{TwiBot-2020}}
& \multicolumn{2}{c}{\textbf{Fox-2023}} \\
\cmidrule(lr){2-3}
\cmidrule(lr){4-5}
\cmidrule(lr){6-7}
& \textbf{Accuracy} & \textbf{Macro-F1}
& \textbf{Accuracy} & \textbf{Macro-F1}
& \textbf{Accuracy} & \textbf{Macro-F1} \\
\midrule
Full
& \textbf{81.63 $\pm$ 0.11} & \textbf{78.27 $\pm$ 0.21}
& \textbf{79.83 $\pm$ 0.18} & \textbf{76.71 $\pm$ 0.26}
& \textbf{96.42 $\pm$ 0.60} & \textbf{94.13 $\pm$ 0.84} \\

5\%
& 80.23 $\pm$ 0.12 & 76.29 $\pm$ 0.15
& 78.28 $\pm$ 0.14 & 74.60 $\pm$ 0.17
& 95.81 $\pm$ 0.08 & 93.08 $\pm$ 0.11 \\

10\%
& 78.74 $\pm$ 0.19 & 74.21 $\pm$ 0.29
& 76.74 $\pm$ 0.22 & 72.51 $\pm$ 0.32
& 94.77 $\pm$ 0.15 & 91.20 $\pm$ 0.29 \\

20\%
& 76.12 $\pm$ 0.24 & 70.19 $\pm$ 0.40
& 74.13 $\pm$ 0.25 & 68.66 $\pm$ 0.40
& 91.99 $\pm$ 0.55 & 85.38 $\pm$ 1.21 \\

40\%
& 70.75 $\pm$ 0.18 & 60.83 $\pm$ 0.36
& 68.73 $\pm$ 0.20 & 59.61 $\pm$ 0.39
& 86.83 $\pm$ 0.46 & 72.04 $\pm$ 1.29 \\

60\%
& 65.80 $\pm$ 0.10 & 50.38 $\pm$ 0.22
& 63.71 $\pm$ 0.11 & 49.51 $\pm$ 0.24
& 82.45 $\pm$ 0.69 & 56.31 $\pm$ 2.98 \\
\bottomrule
\end{tabular}
\end{table*}

\subsection{Analysis of Learned User Representations}
\label{sec:behavior_analysis}

To further investigate whether MGDIL captures meaningful behavioral characteristics beyond classification performance, we analyze the summarized user representations from two perspectives:
(1) whether they preserve discriminative behavioral differences between humans and bots, and
(2) how these behavioral characteristics vary across datasets with different distributions.

Our analysis considers five dimensions derived from users' historical posts: emotional tone, sentiment polarity, linguistic style, content theme, and communicative function.
For clarity, we focus on communicative function and emotional tone in the main paper, while the complete results across all five dimensions are provided in the supplementary material.

\subsubsection{Human-Bot Behavioral Differences}

We first examine whether multi-granularity behavioral summarization captures high-level behavioral characteristics that complement structured profile information.
Overall, humans and bots exhibit distinct behavioral distributions across multiple dimensions, indicating that the generated summaries preserve informative user-level behavioral cues.

As shown in Figure~\ref{fig:function_analysis}(a), communicative function reveals notable differences between humans and bots, although the specific patterns vary across datasets.
In Cresci-2015, humans are more associated with information sharing and opinion expression, whereas bots exhibit stronger tendencies toward self-promotion and presence maintenance.
TwiBot-2020 shows greater overlap between the two classes across several communicative functions.
In contrast, Fox-2023 exhibits a more pronounced separation, with bots concentrated on information sharing, opinion expression, and random statements, while humans remain more associated with self-presentation, social presence, and anecdotal expression.
These findings suggest that communicative function captures meaningful behavioral distinctions, with more pronounced class differences in the newer dataset.

Emotional tone provides complementary behavioral cues.
As shown in Figure~\ref{fig:emotion_analysis}(a), humans are generally more associated with calm or objective expression, whereas bots exhibit stronger tendencies toward emotionally charged or ambiguous communication.
While the differences remain relatively modest in Cresci-2015, bots in TwiBot-2020 show a clearer tendency toward hostile expression.
The strongest divergence appears in Fox-2023, where bots are less associated with calm expression and more frequently exhibit emotional, aggressive, and mixed tones.
Together, these observations indicate that communicative function and emotional tone capture complementary aspects of human-bot behavioral differences, providing informative signals beyond structured profile metadata.

\subsubsection{Cross-Domain Behavioral Shifts}

Although behavioral characteristics provide useful signals for distinguishing humans from bots, their distributions are not necessarily consistent across domains.
We therefore further examine how bot behaviors vary across datasets to investigate the distribution shifts reflected in the summarized representations.

As shown in Figure~\ref{fig:function_analysis}(b), bot communicative functions exhibit substantial cross-dataset variations.
Bots in the earlier Cresci datasets display relatively diverse communicative functions, including social presence and personal-update behaviors.
TwiBot-2020 exhibits a greater emphasis on information sharing while retaining functional diversity.
In contrast, bots in Fox-2023 are predominantly associated with information sharing and opinion expression, whereas personal updates, interpersonal maintenance, and storytelling become less prominent.
These differences suggest a shift in the dominant communicative patterns across datasets, from relatively diverse social interactions toward more concentrated, broadcast-oriented communication.

A similar variation is observed in emotional tone.
As shown in Figure~\ref{fig:emotion_analysis}(b), bots in the Cresci datasets exhibit relatively restrained emotional profiles, characterized by calmer and less hostile expression.
TwiBot-2020 shows an increased prevalence of hostile expression, while Fox-2023 exhibits the strongest tendencies toward emotional, aggressive, and mixed tones.
These cross-dataset differences indicate that emotional characteristics associated with bots can vary substantially across domains, potentially limiting the generalizability of detection methods that rely on fixed behavioral patterns.

Overall, these analyses demonstrate that the summarized behavioral representations capture both discriminative human-bot differences and substantial cross-domain behavioral variations.
The former provides complementary behavioral information for account classification, while the latter highlights the importance of learning domain-invariant representations that preserve discriminative behavioral characteristics under distribution shifts.

\subsection{Robustness to Missing Profile Features}

To evaluate the robustness of MGDIL to incomplete user information, we further randomly mask a proportion of the available profile features on top of the missing values inherently present in each dataset. As shown in Table~\ref{tab:missing_feature_robustness}, masking an additional 5\%--10\% of the features causes only limited performance degradation, indicating that MGDIL is robust to mild information loss. As the masking rate increases, performance declines progressively rather than collapsing abruptly, with substantial reductions in Accuracy and Macro-F1 observed only at masking rates of 40\%--60\%. Overall, these results demonstrate that MGDIL effectively exploits complementary information across different features and maintains relatively stable cross-domain detection performance when part of the user profile information is unavailable.
\begin{figure}[t]
  \centering
  \includegraphics[width=\columnwidth]{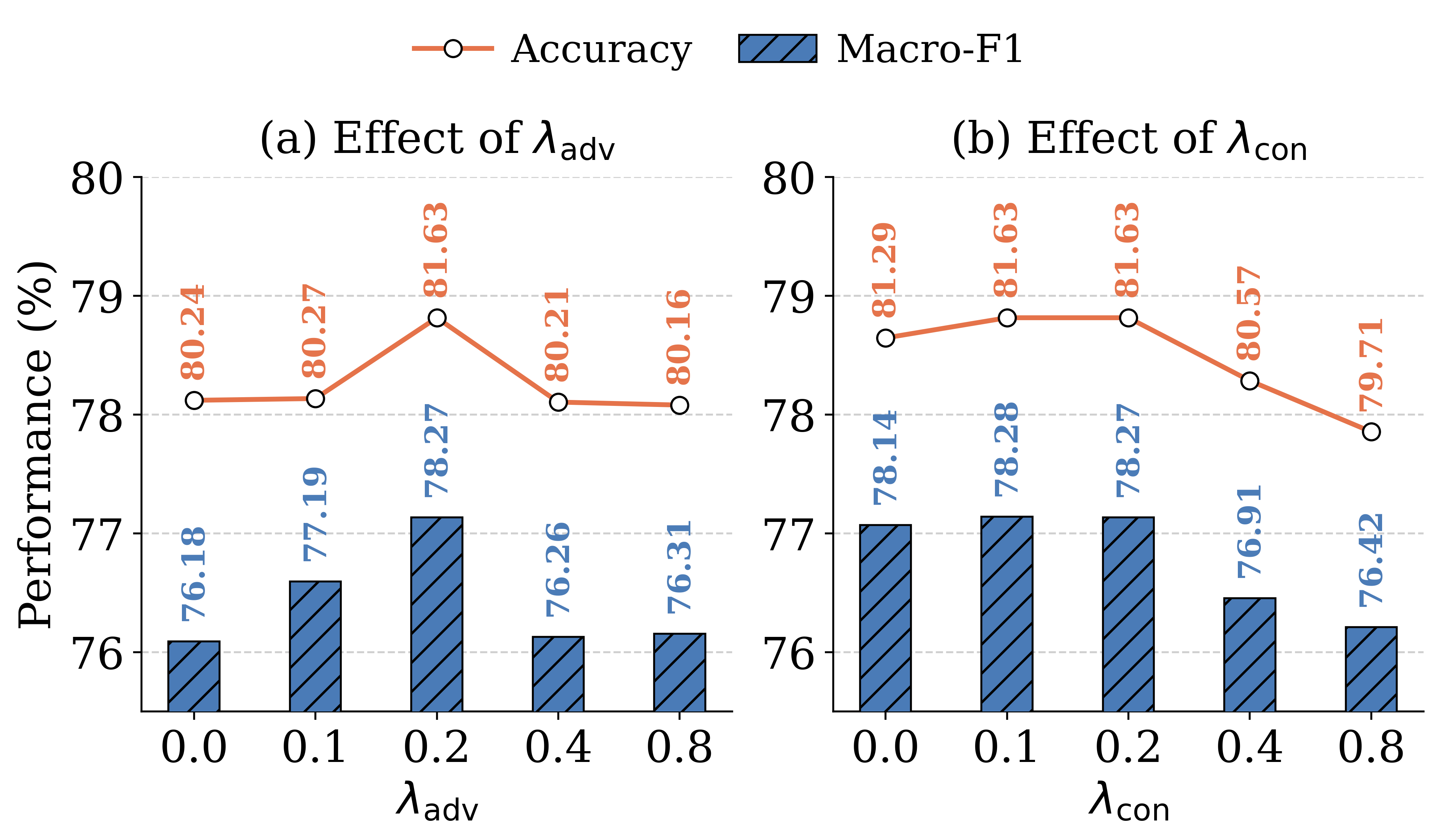}
  \caption{Hyperparameter analysis of $\lambda_{adv}$ and $\lambda_{con}$, where one loss weight is varied while the other is fixed at 0.2.}
  \label{fig:hyperparameter}
\end{figure}
\subsection{Hyperparameter Analysis}

To examine the effect of the two key loss weights, $\lambda_{adv}$ and $\lambda_{con}$, we vary one hyperparameter while fixing the other at 0.2, and report the resulting Accuracy and Macro-F1 in Figure~\ref{fig:hyperparameter}. Overall, both hyperparameters achieve the best performance at relatively moderate values, particularly around 0.1--0.2, whereas excessively small or large weights lead to performance degradation. When varying $\lambda_{adv}$, the model performs best at $\lambda_{adv}=0.2$, achieving the highest Accuracy and Macro-F1. A similar trend is observed for $\lambda_{con}$: performance remains strong when $\lambda_{con}$ is set to 0.1 or 0.2, but declines as the weight increases further. These results indicate that both domain-invariant learning and contrastive learning contribute positively to the model, but their effects need to be properly balanced. Overall, the analysis suggests that setting both loss weights to 0.2 yields stable and effective performance, motivating our final choice of $\lambda_{adv}=\lambda_{con}=0.2$.

\section{Conclusions}

We presented MGDIL, a framework for generalizable social bot detection that integrates multi-granularity summarization with domain-invariant representation learning.
The summarization module converts heterogeneous profile attributes and historical posts into a unified textual representation, while task-oriented instruction tuning, domain-adversarial learning, and cross-domain contrastive learning jointly reduce dependence on source-domain-specific patterns while preserving bot-human discriminability.
Experiments on 15 datasets demonstrate that MGDIL improves over representative baseline families under the considered cross-dataset evaluation settings.
The embedding-transfer study further shows that the learned representations provide effective node features for graph-based detectors when relational information is available.
Future work will investigate finer-grained domain definitions, robustness under controlled information-missing settings, and the computational trade-offs of LLM-based summarization.

\bibliographystyle{IEEEtran}
\bibliography{references}

\end{document}